\documentclass[manuscript]{aastex}
\usepackage{graphicx}
\usepackage{subfigure}

\shorttitle{The Twist of the Draped Interstellar Magnetic Field Ahead of the Heliopause:}
\shortauthors{Opher, Drake, Swisdak, Zieger and Gabor}

\begin{document}

\title{The Twist of the Draped Interstellar Magnetic Field Ahead of the Heliopause: A Magnetic Reconnection Driven Rotational Discontinuity}

\author{M. Opher\altaffilmark{1,4}}
\affil{Astronomy Department, Boston University, Boston, MA 02215}
\email{mopher@bu.edu}

\author{J. F. Drake\altaffilmark{2}}
\affil{Department of Physics, the Institute for Physical Science and Technology and the Joint Space Science Institute, University of Maryland, College Park, MD}

\author{M. Swisdak\altaffilmark{3}}
\affil{Insitute for Research in Electronics and Applied Physics, University of Maryland, College Park, MD}
 
\author{B. Zieger\altaffilmark{4}}
\affil{Center for Space Physics, Boston University, Boston, MA 02215}

\author{G. Toth\altaffilmark{5}}
\affil{University of Michigan, Ann Arbor, MI}

\begin{abstract}
Based on the difference between the orientation of the interstellar $B_{ISM}$ and the solar magnetic fields, there was an expectation that the magnetic field direction would rotate dramatically across the heliopause (HP). However, the Voyager 1 spacecraft measured very 
little rotation across the HP. Previously we showed that the $B_{ISM}$ twists as it approaches the HP and acquires a strong
 T component (East-West). Here we establish that reconnection in the eastern flank of the heliosphere is responsible for the twist.
 On the eastern flank the solar magnetic field has twisted into the positive N direction and reconnects with the Southward pointing component of the $B_{ISM}$. Reconnection drives a rotational
discontinuity (RD) that twists the $B_{ISM}$ into the -T direction and propagates upstream in the interstellar medium towards the nose.
The consequence is that the N component of $B_{ISM}$ is reduced in a finite width band upstream of the HP. Voyager 1 currently
measures angles ($\delta=sin^{-1}(B_{N}/B)$) close to solar values. We present 
MHD simulations to support this scenario, suppressing reconnection in the nose region while allowing it in the
flanks, consistent with recent ideas about reconnection suppression from diamagnetic drifts. The jump in plasma $\beta$ (the plasma to magnetic pressure) across the nose of HP is much greater than in the flanks because the heliosheath $\beta$ is greater there than in the flanks. Large-scale reconnection is therefore suppressed in the nose but not at the flanks. 
Simulation data suggest that $B_{ISM}$ will return to its pristine value $10-15~AU$ past the HP.
\end{abstract}


\section{Introduction}
On August 25, 2012, the solar wind particles (with energy of 10's of keVs) measured by Voyager 1 dropped to noise level as it crossed into
the interstellar medium (Stone et al. 2013). Based on the measured heliospheric asymmetries there was the prediction (Izmodenov et
al. 2009; Opher et al. 2006, 2009, Pogorelov et al. 2007) that the direction of interstellar magnetic field ($B_{ISM}$) would be highly
inclined with respect to the east-west direction of solar magnetic field. Therefore, the expectation was of a dramatic rotation of the
magnetic field direction at the heliopause. However, when Voyager 1 crossed the heliopause (HP), observations (Burlaga et al. 2013) revealed
that the magnetic field had almost no change in the direction or magnitude. These observations sparked suggestions that the conditions
seen by Voyager 1 could be explained by temporal instabilities or flux transfer events (Krimigis et al. 2013; Florinski et al. 2015;
Schwadron \& McComas 2013, Borovikov \& Pogorelov 2014).  Now that Voyager 1 has been in the interstellar medium (ISM) more than three years
(Burlaga et al. 2016) (or $16~AU$ past the heliopause) and is observing nearly constant orientation and magnitude of $B_{ISM}$,
temporal processes can be ruled out.

Previously we suggested (Opher \& Drake 2013) that the draping $B_{ISM}$ around the heliopause is strongly affected by the solar
magnetic field although the physical mechanism for such behavior was not understood. We showed in magnetohydrodynamic (MHD)
simulations that as $B_{ISM}$ approaches the heliopause, it twists and acquires an east-west component, which did not occur if the solar
magnetic field was not present. It was suggested that the draping of $B_{ISM}$ around the heliopause could explain the twist of $B_{ISM}$
and the Voyager 1 observations (Isenberg et al. 2015, Grygorczuk et al., 2014). Here, we present an alternative idea, that magnetic
reconnection at the eastern flank of the heliosphere is responsible for most of the twist of the interstellar magnetic field outside of
the HP along the Voyager 1 trajectory.

In global MHD simulations (Opher et al 2015, Pogorelov et al 2015, Izmodenov and Alexashov 2015) the solar magnetic field in the heliosheath twists in 
the negative N direction on the eastern flank, opposing that of $B_{ISM}$, which has a positive N component. We show that physical
conditions allow reconnection to proceed between $B_{ISM}$ and the solar magnetic field in the eastern flank while being suppressed at
the nose. A rotational discontinuity (RD) forms at the reconnection site on the eastern flank and propagates upstream of the nose. The
consequence is that there is a finite band of reconnected field lines outside of the HP where the magnetic field is mostly in the T
direction and the elevation angle $\delta=sin^{-1}(B_{N}/B)$ is close to the solar value, consistent with the Voyager 1 measurements. Once
Voyager 1 crosses the domain where the RD of the reconnected field has propagated upstream it will measure larger values of $B_{N}$ and the
associated angle $\delta$. Here the T and N direction refer to the RTN coordinate system
that is the Cartesian system centered in the spacecraft. R is radially outward from the Sun, T is in the plane of the solar equator and is
positive in the direction of solar rotation, and N completes a right-handed system. 

\section{MHD Model}
We use the same model as in Opher et al. (2015), a multi-fluid magnetohydrodynamic (MHD) 3D model with
one ionized fluid and four fluids for the neutral H component (Opher et al. 2009) based on the 3D multi-fluid MHD code BATS-R-US 
with adaptive mesh refinement (Toth et al. 2012). The multi-fluid approach for the neutrals captures the main features of the kinetic model 
(Izmodenov et al. 2009). 

The inner boundary of our domain is a sphere at $30$AU and the outer boundary is at $x = \pm1500$AU, $y = \pm 1500$AU, 
$z = \pm 1500$AU. Parameters of the solar wind at the inner boundary at $30$AU are: $v_{SW} = 417$ km/s, 
$n_{SW} = 8.74 \times 10^{-3} cm^{-3}$, $T_{SW} = 1.087 \times 10^{5}$K (OMNI solar data, http://omniweb.gsfc.nasa.gov/). The Mach number of the solar wind is
$7.5$ and is therefore super-fast-magnetosonic. Therefore all the flow parameters can be specified at this boundary. The magnetic field is
given by the Parker spiral magnetic field (Parker 1958).

We assume that the magnetic axis is aligned with the solar rotation axis. The solar wind flow at the inner boundary is assumed to be
spherically symmetric. The coordinate system is such that the z-axis is parallel to the solar rotation axis, the x-axis is $5^{\circ}$
above the direction of interstellar flow with y completing the right-handed coordinate system.

We use a monopole configuration for the solar magnetic field. This description while capturing the topology of the field line doesn't
capture its change of polarity with solar cycle or across the heliospheric current sheet. This choice, however, minimizes artificial
reconnection effects, especially in the heliospheric current sheet. Such procedure was used by other groups (e.g., Izmodenov et
al. 2015, Zirnstein et al. 2016). We chose the solar field polarity that corresponds to solar cycle 24, with the azimuthal angle $\lambda$
(between the radial and T directions in heliospheric coordinates) $270^{\circ}$ in the north and south. The interstellar magnetic field
has a T component of $270^{\circ}$ as detected by Voyager 1. Such configuration minimizes reconnection at the nose (Figure 1a).

Here we show results from two different simulations with different orientations for the $B_{ISM}$. Model A has the $B_{ISM}$ in
the hydrogen deflection plane (HDP) ($-34.7^{\circ}$ and $57.9^{\circ}$ in ecliptic longitude and latitude, respectively) consistent with the measurements of deflection of He atoms with respect to the H atoms (Lallement et
al. 2005; 2010). Model B is the one used in works that constrain the orientation of $B_{ISM}$ based on the circularity of the IBEX ribbon
and the ribbon location (Zirnstein et al. 2016) ($-34.62^{\circ}$ and $47.3^{\circ}$ in ecliptic longitude and latitude, respectively) . The specific orientation of $B_{ISM}$ for the present paper is not important since
the solar wind conditions are idealized so the exact shape of the heliosphere is not important. The main point of this paper depends on
the fact that the interstellar field is highly inclined to the east-west direction, which is true for both cases.

Model A has $v_{ISM} = 26.4$km/s, $n_{ISM} = 0.06 cm^{-3}$, $T_{ISM}= 6519$K. Model B has $v_{ISM} = 25.4$km/s, $n_{ISM} = 0.0925 cm^{-3}$, $T_{ISM}= 7500$K. The magnitude of $B_{ISM}$ is $4.4$nT (model A) and $2.93$ nT (model B). The number density of H atoms in the interstellar medium is $n_{H} = 0.18 ~cm^{-3}$ (model A) and $n_{H} = 0.155~cm^{-3}$ (model B). 

Models A and B were run to 260,000 time steps which corresponds to 231 years with $9.11\times 10^{7}$ cells with resolution equivalent to the one
used in (Opher et al. 2016) with minimum grid resolution of $0.37$ AU near the HP and $93.75$ AU farther out. For Model B we then
used Adaptive Mesh Refinement to create a high-resolution grid around the HP ($0.36$ AU at the HP and $0.18$ AU along the Voyager 1 trajectory resulting in 
$2.4 \times 10^{9}$ cells. The HP is defined as temperature iso-surface with $T=2.683 \times10^{5}$K (Figure 1a)

\section{Reconnection and Transport and Convection of the Interstellar Magnetic Field}

We recently found that the magnetic tension of the solar magnetic field plays a crucial role in organizing the solar wind (Opher et
al. 2015; Drake et al. 2015) in the heliosheath into two jet-like structures. The heliosphere then has a ``croissant-like shape'' where
the distance to the heliopause downtail is almost the same as towards the nose. This new view is vastly different from the standard picture
of the heliosphere as a comet-shape like structure with the tail extending for 1000's of AUs.

However, we argue here that the detailed of the shape of the heliosphere far downstream is less important than the orientation of
the solar magnetic field as it convects downstream, which is the same in all of the global models.  As shown in Opher et al. 2015 and Opher
et al. 2016 the solar magnetic field (shown in red - Figure 1) as it convects down the tail maintains a ``tube-like'' topology that is bent
due to the flow of interstellar medium (ISM). The bent tube topology twists the solar magnetic field, which is initially in the (east-west) T
direction at the nose, towards the N direction in the flanks (Figure 1a). This magnetic geometry of the solar magnetic
field is also seen in global MHD simulations that display an extended tail (Pogorelov et al. 2015). Therefore, regardless of whether the
tail is split or not, the solar magnetic field is mostly in the N direction in the flanks. A $B_{ISM}$ that has positive T and negative R
and N components first contacts the heliosphere in the south-east hemisphere (where east and west refers to a view from the ISM towards
the Sun) (Figure 1a). On the eastern flank it encounters the solar magnetic field that is mostly oriented in the positive N direction (for
the chosen polarity of the solar magnetic field). The N components of the solar and interstellar field reconnect (Figure 1a), leaving a
remnant positive guide-field component $B_{T}$.

Once reconnection occurs it creates a pair of field lines, one of which is the solar magnetic field that now is open to the interstellar
medium (Figure 1b).  A second reconnection can happen with another $B_{ISM}$ field line (Figure 1c) that leads to a reconnected magnetic field
that is open at both ends into the ISM (Figure 1d). The field lines then get convected and stretched towards the northern and southern
poles (Figure 1d- right panel). As the ISM gets convected towards the heliosphere new ISM field lines wrap around the heliosphere and the
cycle repeats.

\begin{figure}
%[htbp]
\centering
%\\
\begin{subfigure}{(a)}
    \includegraphics[width=0.23\textwidth]{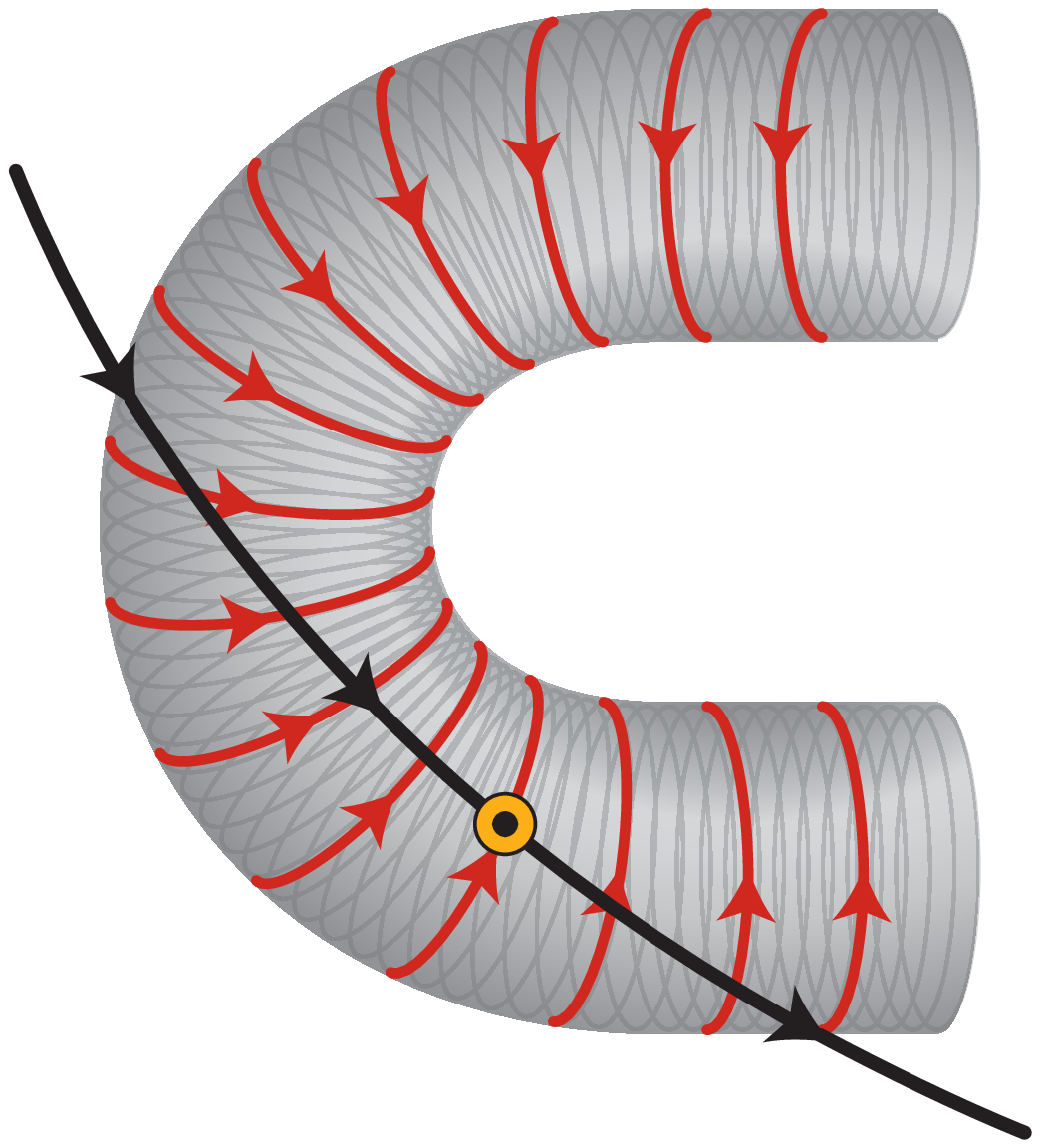}  
    \includegraphics[width=0.25\textwidth]{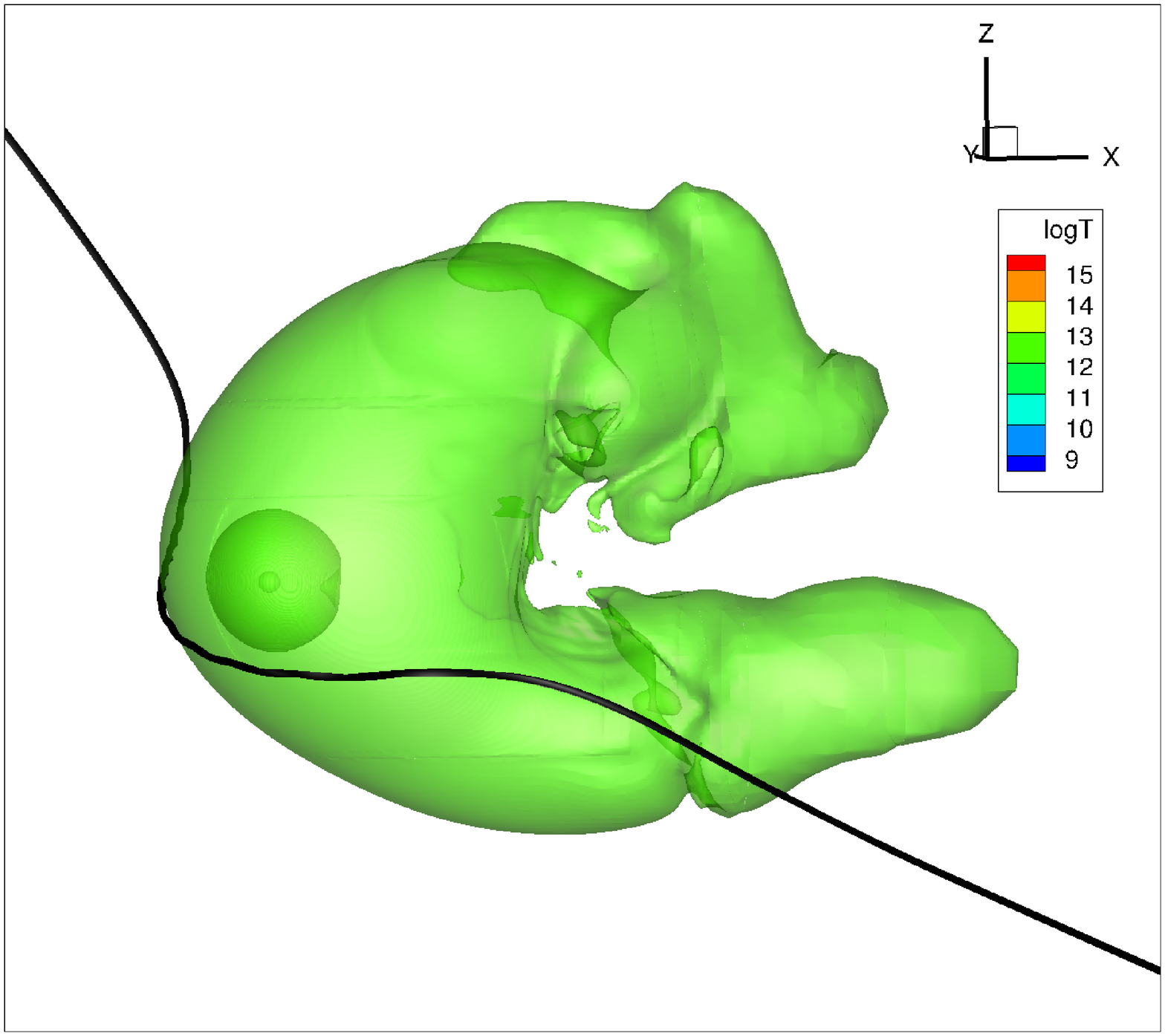}   
\end{subfigure} %
\\
\begin{subfigure}{(b)}
    \includegraphics[width=0.23\textwidth]{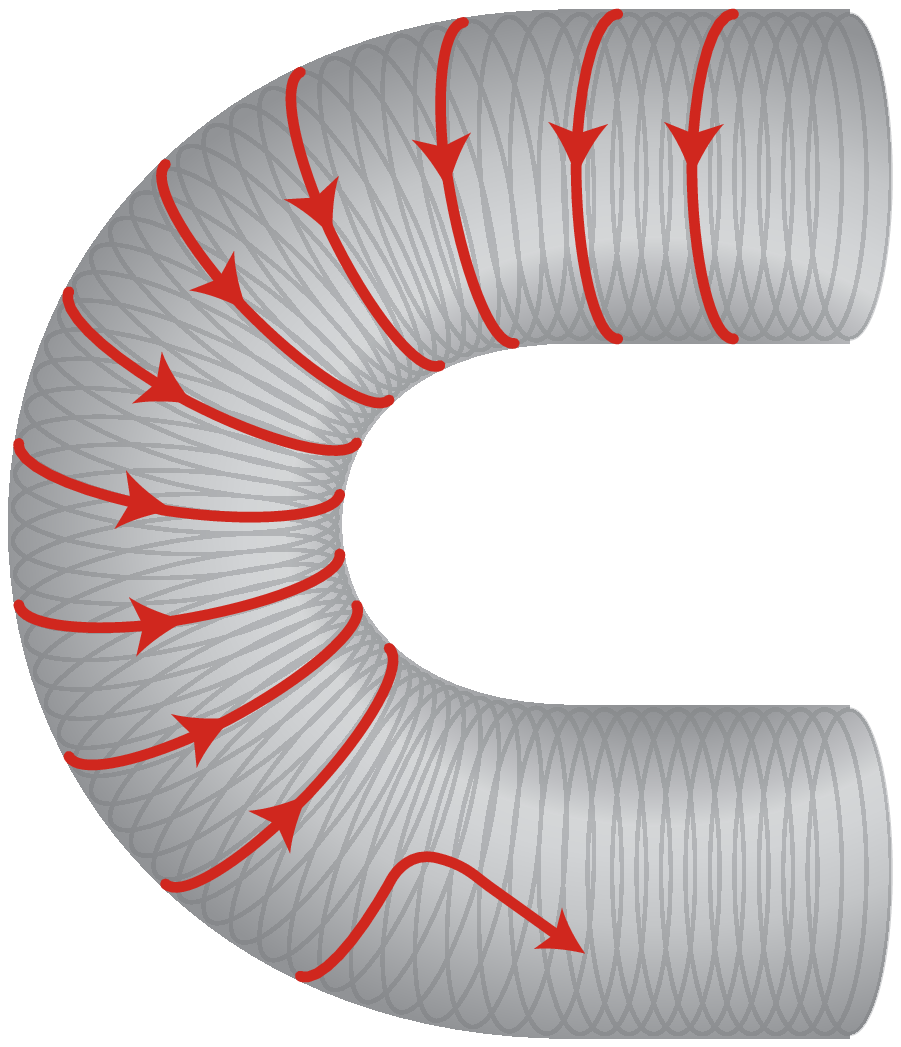}  
    \includegraphics[width=0.25\textwidth]{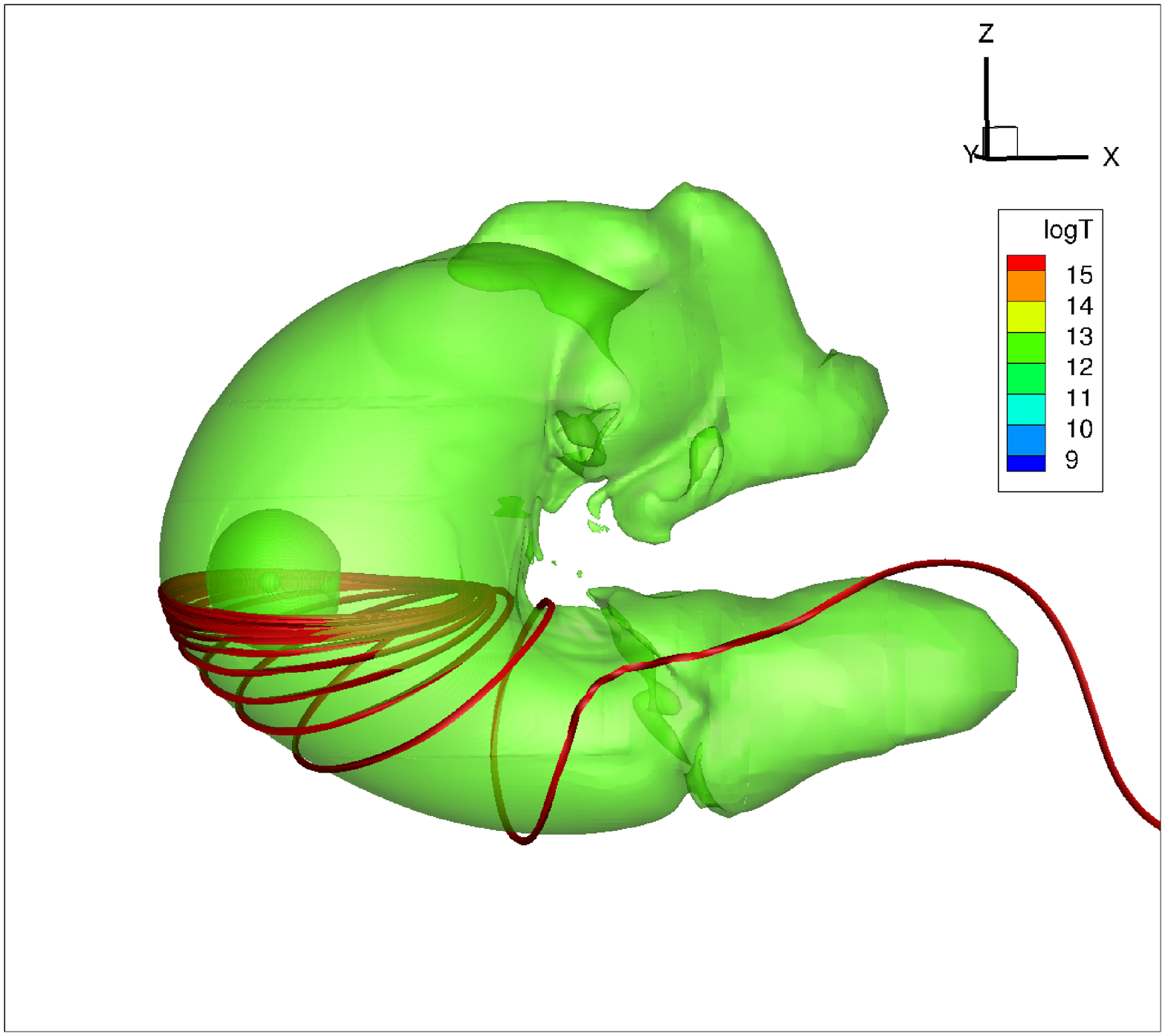}    
\end{subfigure} %%
\\
\begin{subfigure}{(c)}
    \includegraphics[width=0.23\textwidth]{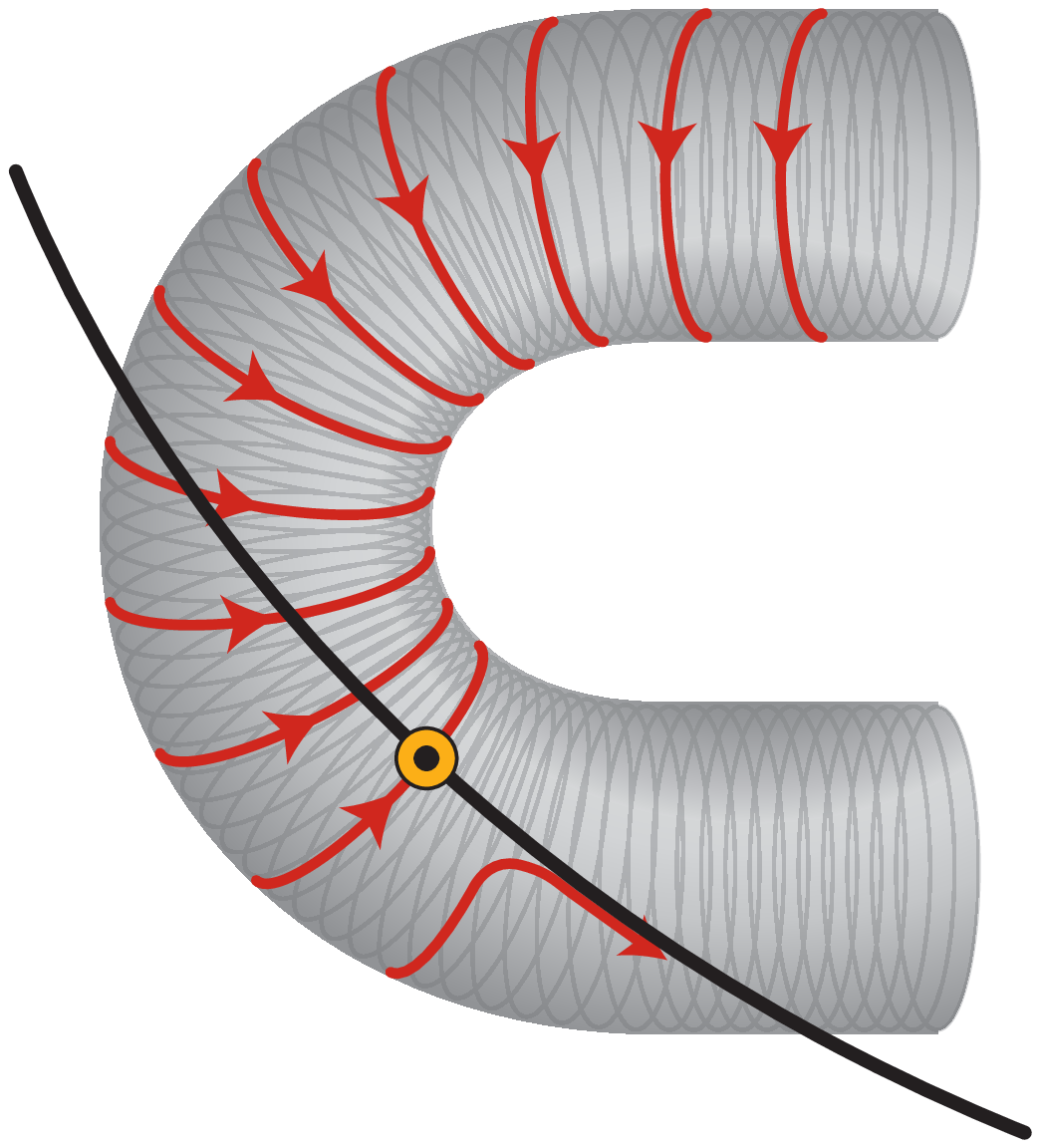}          
    \includegraphics[width=0.25\textwidth]{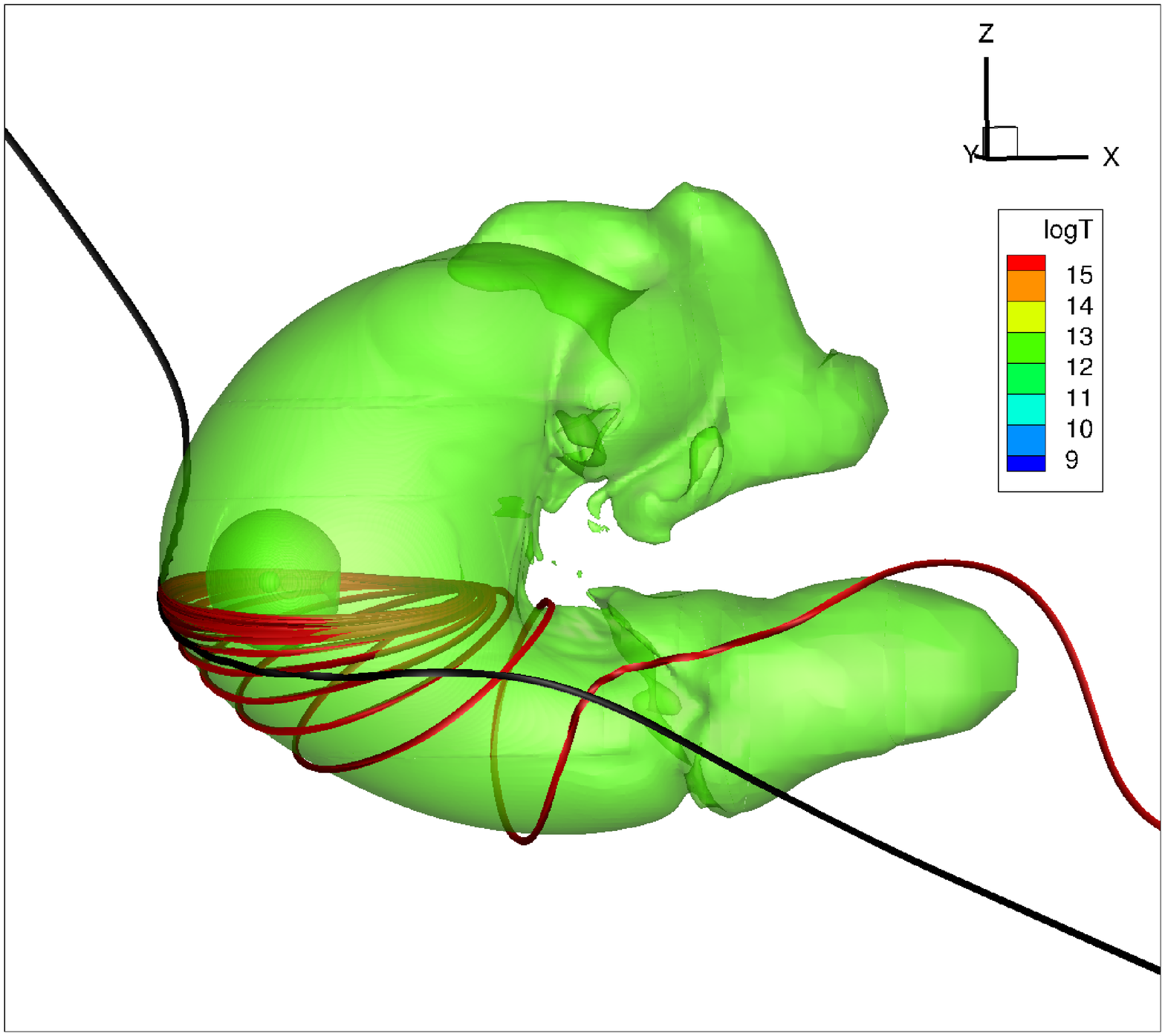}         
\end{subfigure} %
\\
\begin{subfigure}{(d)}
    \includegraphics[width=0.23\textwidth]{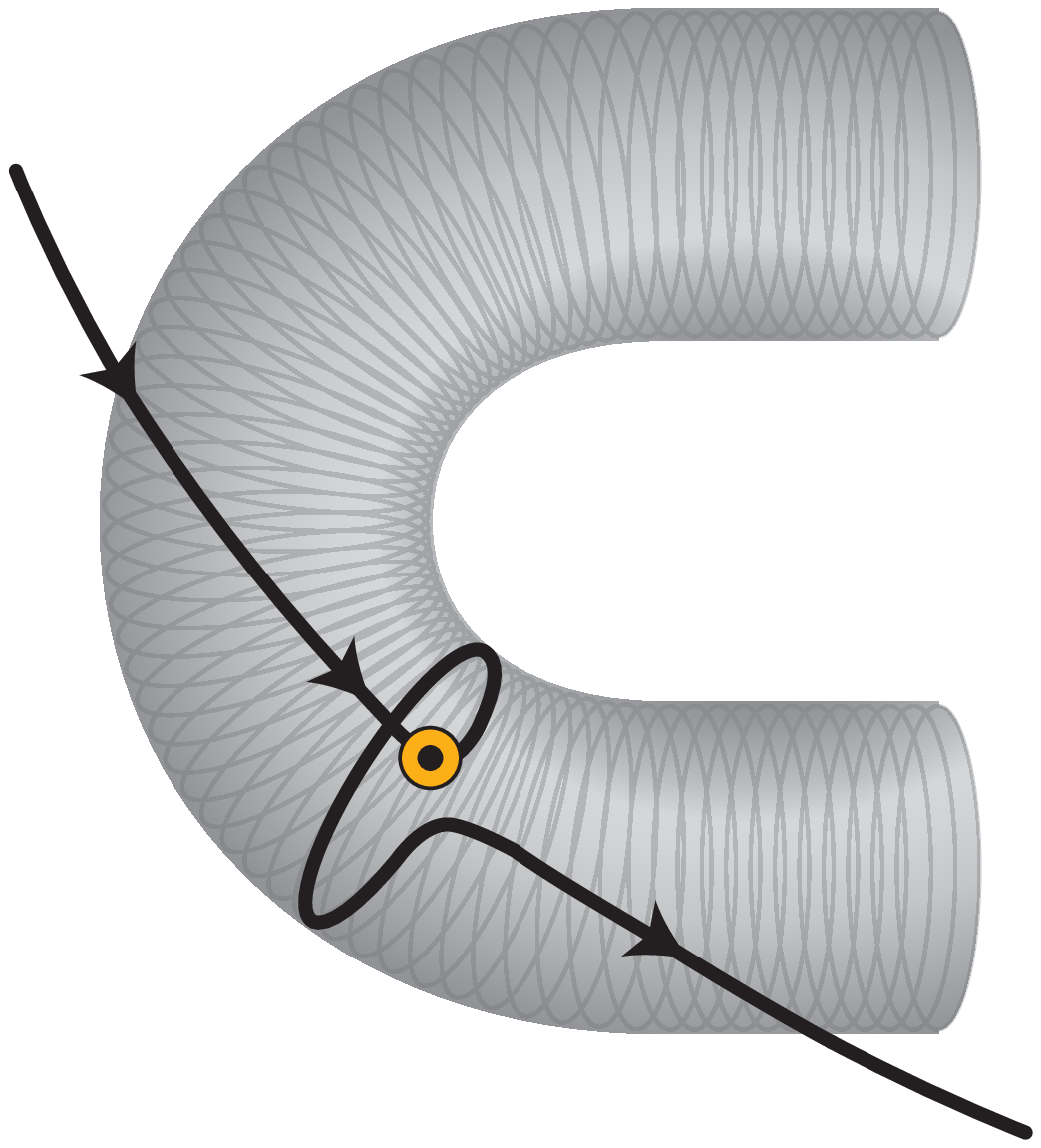}    
    \includegraphics[width=0.25\textwidth]{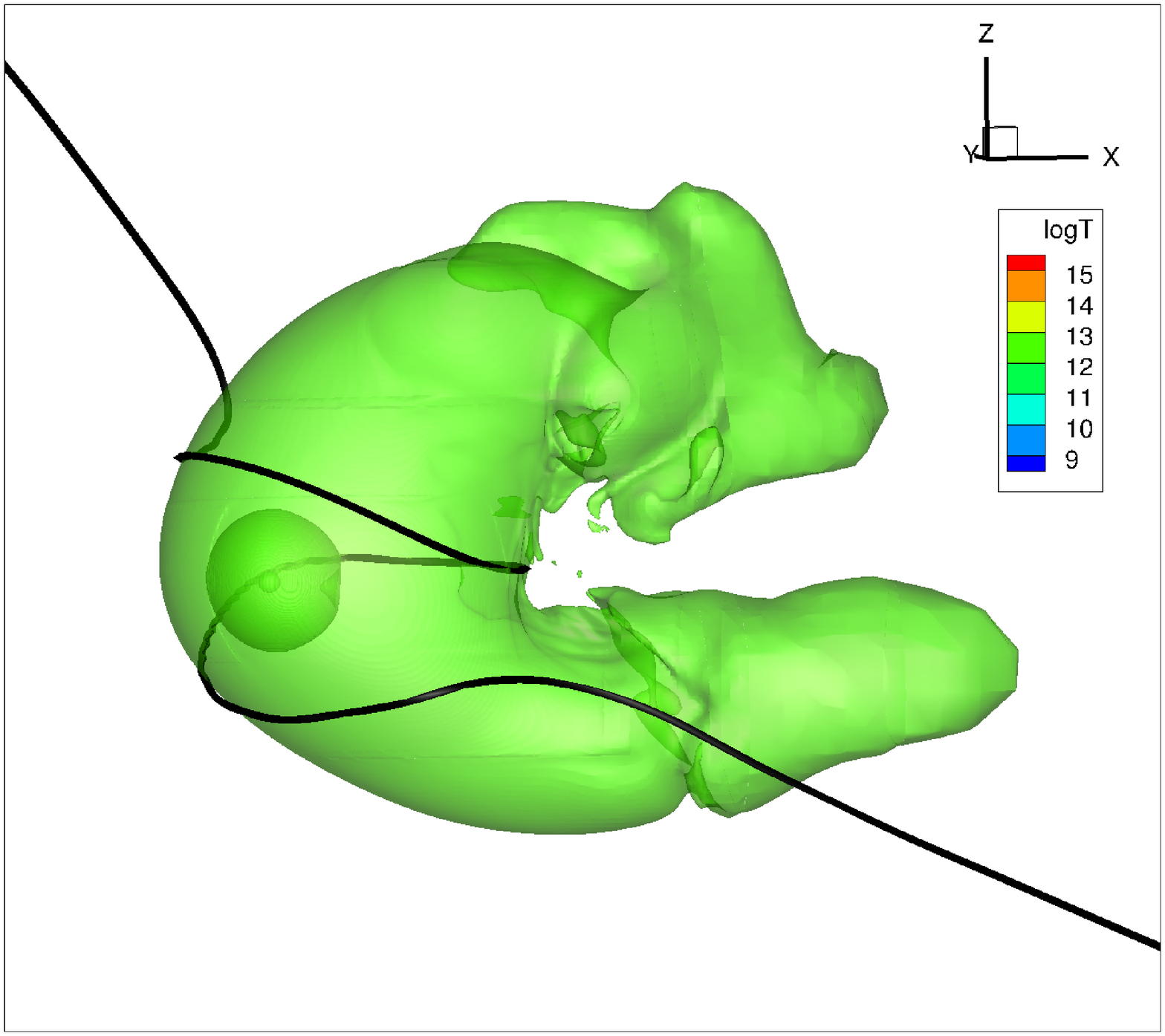}      
\end{subfigure} %     
\caption{The pattern of convection and transport of the interstellar magnetic field lines around the heliosphere. The left column show a series of cartoons while the right hand column shows magnetic field lines taken from the 3D MHD simulation that exemplify each cartoon. The solar magnetic field is shown in red while the interstellar magnetic field is shown in black. The heliopause is shown in the 3D MHD simulation by an iso-surface of temperature $T=2.683\times 10^{5} K$ in green and in gray in the cartoon (left column). The yellow circle indicates the reconnection site. The MHD model used here is model A.}
\label{figure1}
\end{figure}
 
Figure 2 shows the locations of reconnection in the MHD simulation, traced by $\vec{\nabla} \times \vec{B}/\mid B \mid$. One can see that for the
polarity chosen for the solar magnetic field, reconnection is suppressed at the nose but proceeds at the flanks. The solar magnetic
field becomes turbulent down the flanks (see also Pogorelov et al. 2015) and reconnection becomes widespread.

\begin{figure}
\centering
\includegraphics[width=0.4\textwidth]{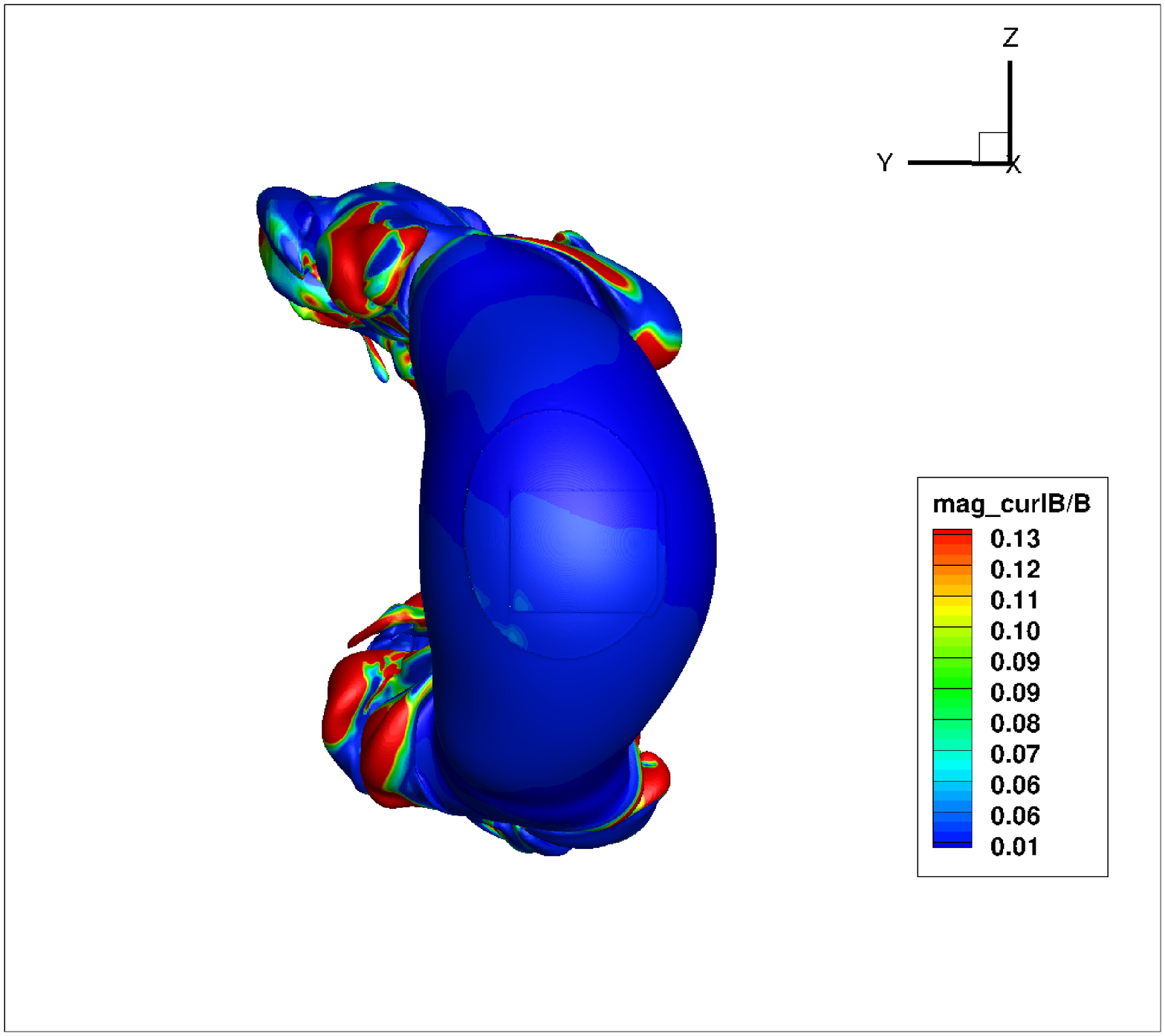}  
\includegraphics[width=0.4\textwidth]{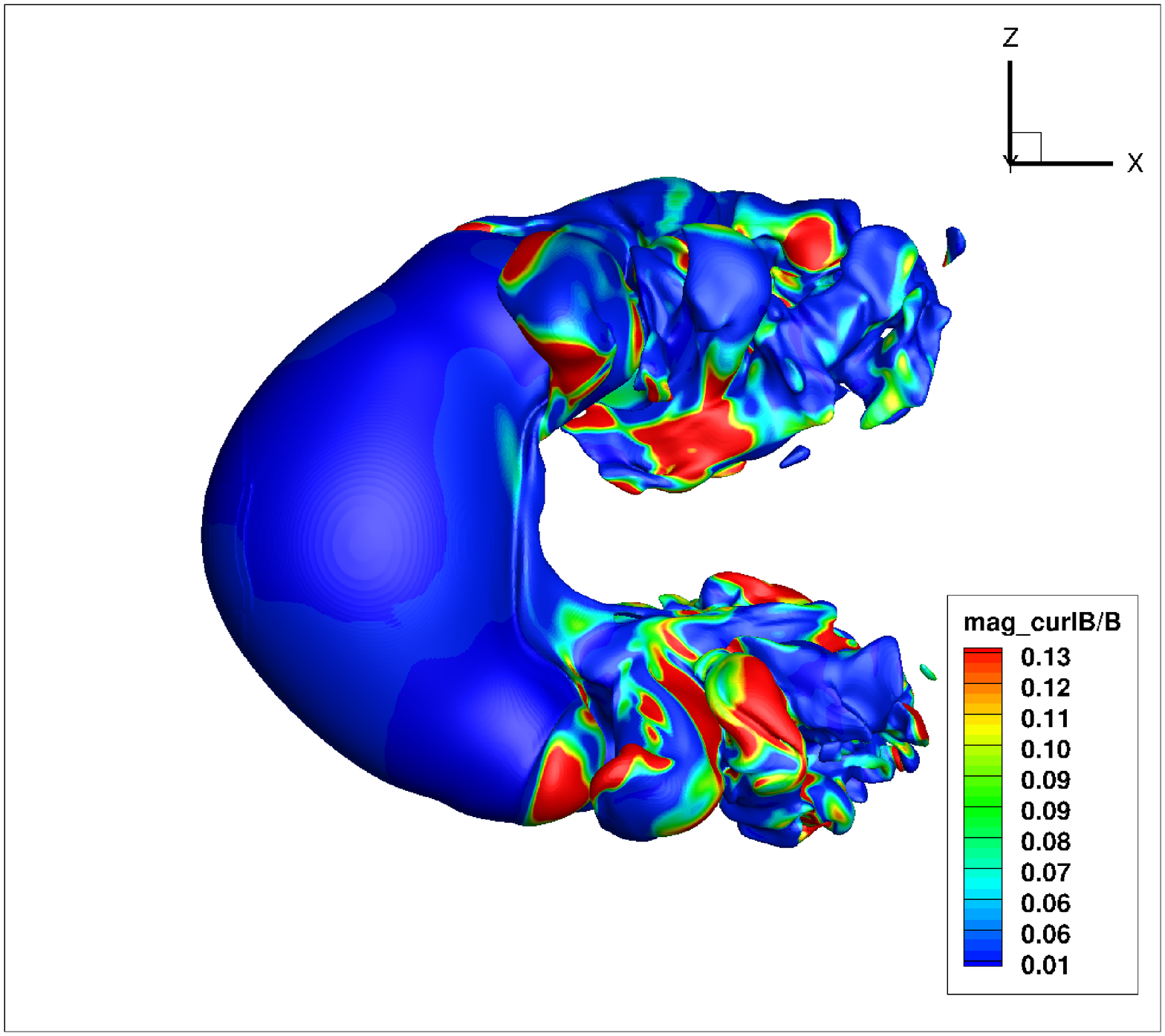}  
\caption{Reconnection spots in the heliosphere. Contours of  $(\vec{\nabla} \times \vec{B})/\mid B \mid$ are plotted on the Heliopause (captured by an iso-surface of $lnT=12.5$ ($T=2.68 \times 10^{5}~K$)). In regions where reconnection is occurring we expect $(\vec{\nabla} \times \vec{B})/\mid B \mid$ to be large. One can see that reconnection is suppressed in the nose (a); (b) while its taking place in the flanks - east view. The MHD model used here is model A.}
\label{figure2}
\end{figure}

For the simulation presented in Fig. 1 the polarity of the solar magnetic field is such that reconnection is not possible at the nose
and can take place in the eastern flank. We argue later that this simulation actually captures the physical conditions expected in the
heliosphere.

\section{Rotational Discontinuity in the Eastern Flanks}

Reconnection in the eastern-flank between the normal component of the $B_{ISM}$ and the solar magnetic field that is oriented mainly in the
N direction creates a pair of rotational discontinuities (RD) in the reconnection exhaust. The left panel of Figure 3 shows the
configuration in the N-R plane. At the RD the normal component of $B_{ISM}$, $B_{N}$ goes to zero (or to some reduced value) leaving a
dominant $B_{T}$. The RD propagates towards the nose of the heliosphere as an Alfv\'en kink. The reduction of $B_{N}$ by the RD
eliminates the rotation of magnetic field across the heliopause and in a finite domain outside the HP whose width is controlled by the tilt
of $B_{ISM}$, transit time of the RD from the east flank to the nose and the radial flow of the ISM outside of the HP. The $B_{ISM}$ will
rotate back to its interstellar orientation outside of the region accessible to the RD. The right hand side panel shows the same view in
the the N-T plane. The kink in magnetic field (dashed line) with reduced $B_{N}$ propagates upstream toward the nose along $B_{ISM}$. 

\begin{figure}[htbp]
\centering
\includegraphics[width=0.45\textwidth]{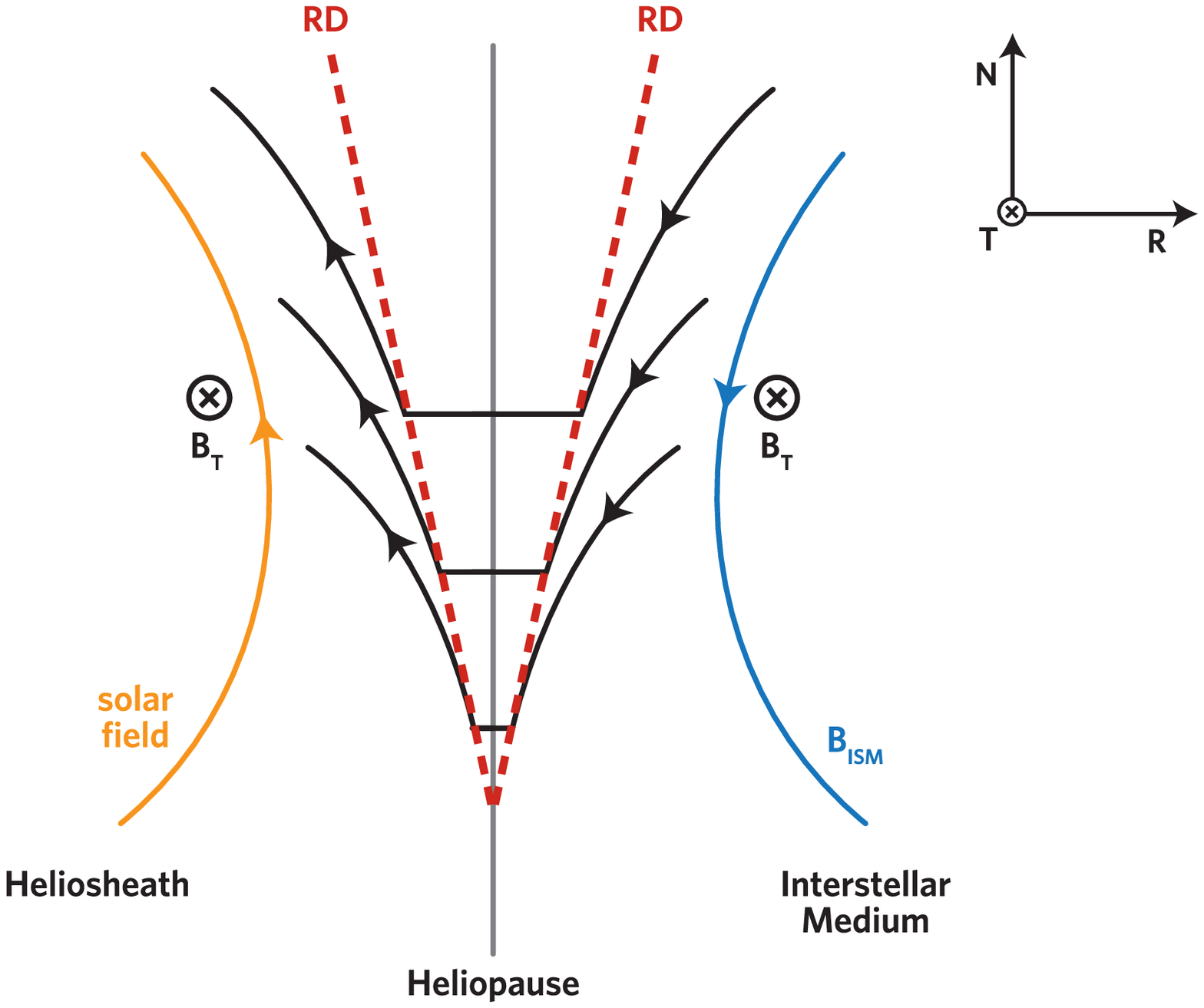}  
\includegraphics[width=0.45\textwidth]{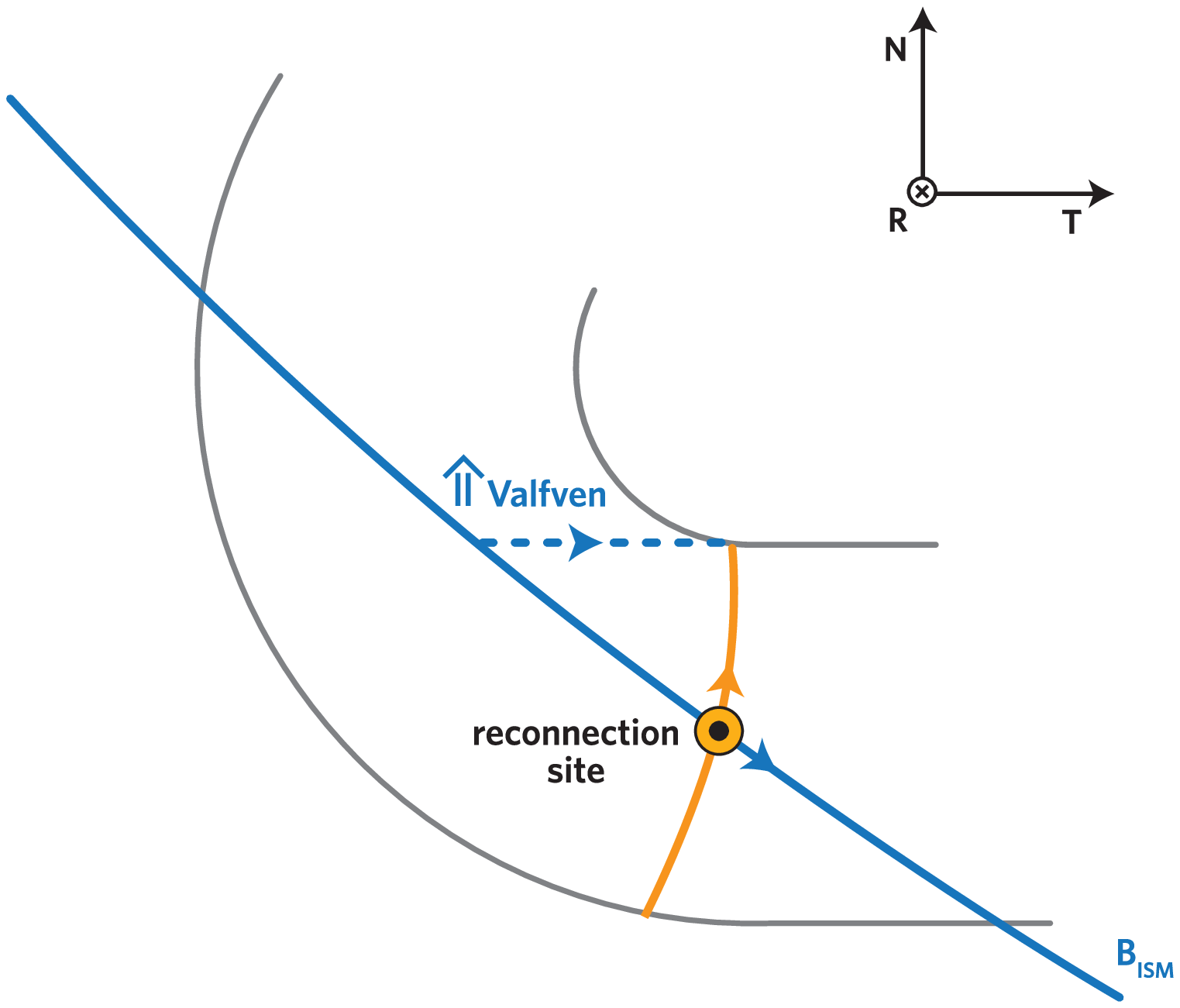}  
\caption{The left panel shows a cartoon with a view of the reconnection in the east-flank in the N-R plane. The $B_{ISM}$ here is chosen to have 
mostly a N and T components. The solar magnetic field in the eastern flank is predominantly in the positive N direction. A pair of RDs form at the boundary 
of the reconnection exhaust (red dashed lines) that rotates $B_{ISM}$ from the T-N direction to the T direction. The RD propagates as as 
Alfven kink along $B_{ISM}$ (the N-T plane in the right panel). The kink in magnetic field (dashed line) reduces $B_{N}$ upstream. }
\label{figure3}
\end{figure}

In the flanks of the HS the plasma $\beta = p_{T}/p_{B}$ (where $p_{T}$ is the thermal pressure and $p_{B}$ is the magnetic pressure),
which is dominated by shock-heated pickup ions, should be much lower than in the nose since the interstellar neutral wind, which is
dominantly in the T direction (Figure 3b), does not penetrate into the high-velocity solar wind upstream of the TS. The generation of pickup
ions should therefore be much less efficient in the flanks. The consequence is that because of the reduced jump in $\beta$ across the
HP in the flanks, reconnection will be much more robust there than in the nose (Swisdak et al 2010).

In the ISM the Alfv\'en speed ($\sim 120$ km/s) is higher than the interstellar flows around the heliopause ($\sim 30$ km/s) so the RD
will be able to propagate along $B_{ISM}$ into the nose region outside of the HP. Also because the plasma $\beta$ is low the reconnected
field lines affected by flank reconnection will contain only interstellar plasma in the nose region outside of the HP.

Once the RD reaches the nose it rotates the magnetic field lines ahead of the HP into the T direction (Fig. 1). To illustrate the behavior of
the magnetic field as Voyager 1 moves from the HP into the interstellar medium crossing the RD we present a cut along a
trajectory approximately around the Voyager 1 latitude from model B (Figure 4). The exact location is not important since our heliosphere
is idealized (e.g., constant solar wind speed) so we do not expect that our MHD model will quantitatively reproduce the real shape of the
heliosphere. One can see that outside of the HP along the R direction the interstellar magnetic field undergoes a gradual rotation in which
the angle $\delta$ (and $B_N$) increases. The right black line in Fig. 4 is the innermost interstellar field line that reconnected on the eastern
flank with an RD that is able to reach the nose region. All field line outside of the gray area are causally disconnected (by Alfv\'en waves) from the eastern flank.

\begin{figure}[htbp]
\centering
\includegraphics[width=0.3\textwidth]{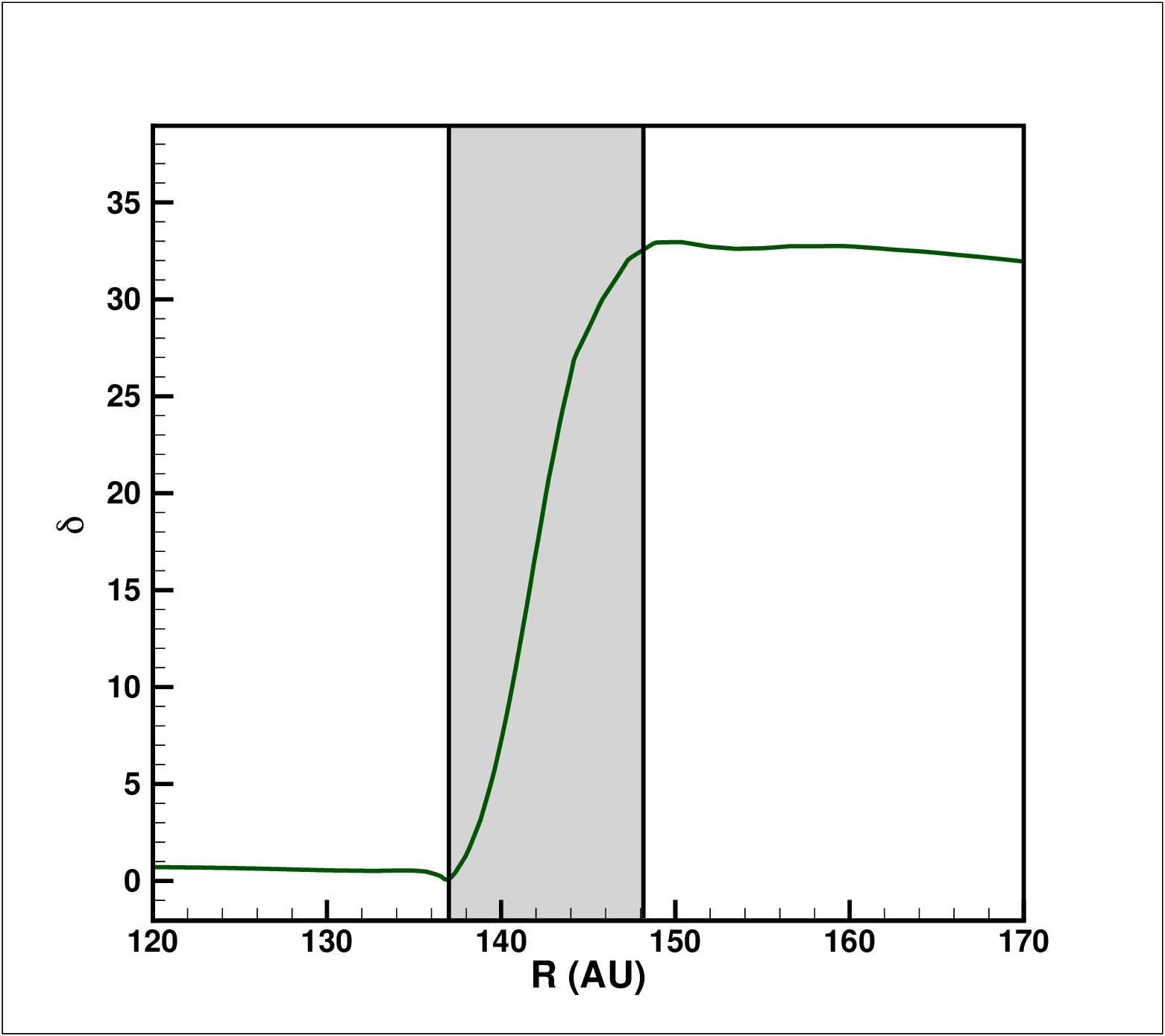}  
\includegraphics[width=0.3\textwidth]{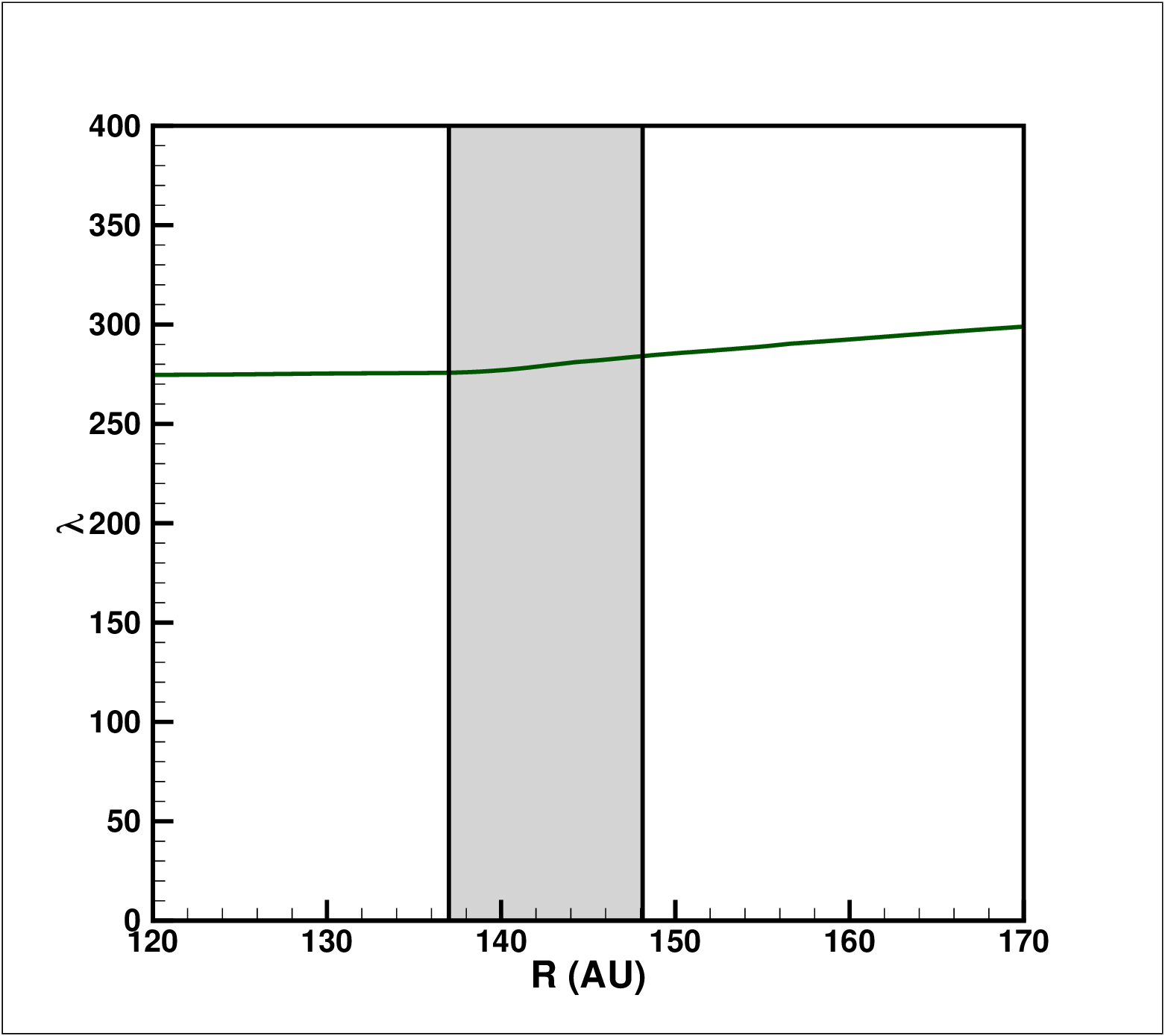}  
\includegraphics[width=0.3\textwidth]{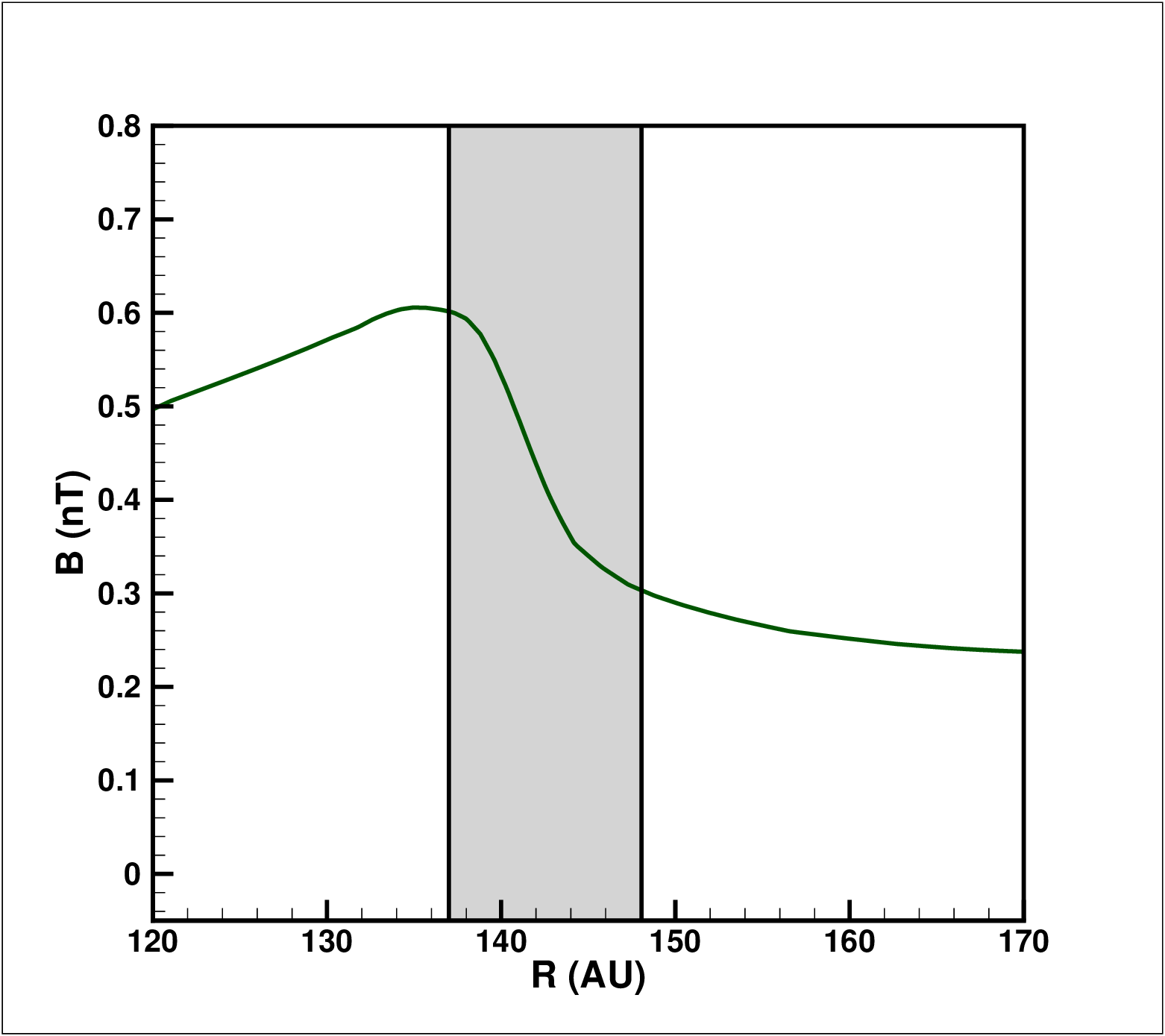}          
\caption{Cut along a trajectory approximately around the Voyager 1 trajectory. Panel (a) shows the angle $\delta= \sin^{-1}(B_{N}/B)$;
  panel (b) $\lambda = \tan^{-1}(B_{T}/B_{R})$ and (c) the magnitude of the magnetic field; where the $RTN$ coordinate system is the local
  Cartesian system centered at the spacecraft. R is radially outward from the Sun, T is in the plane of the solar equator and is positive in the
  direction of solar rotation, and N completes a right-handed system. The black line indicates the heliopause and the dashed line is the 
  innermost interstellar field line that is causally connected (by Alfv\'en waves) to the solar magnetic field at the eastern
  flank. The MHD model used here is model B.}
\label{figure4}
\end{figure}

Figure 5 shows the magnetic field lines ahead of the heliopause. One can see that $B_{ISM}$ retains a solar-like orientation (red lines that have reconnected with the solar magnetic field on the flanks) between the HP and the green, unreconnected field lines.

\begin{figure}[htbp]
\centering
\includegraphics[width=0.35\textwidth]{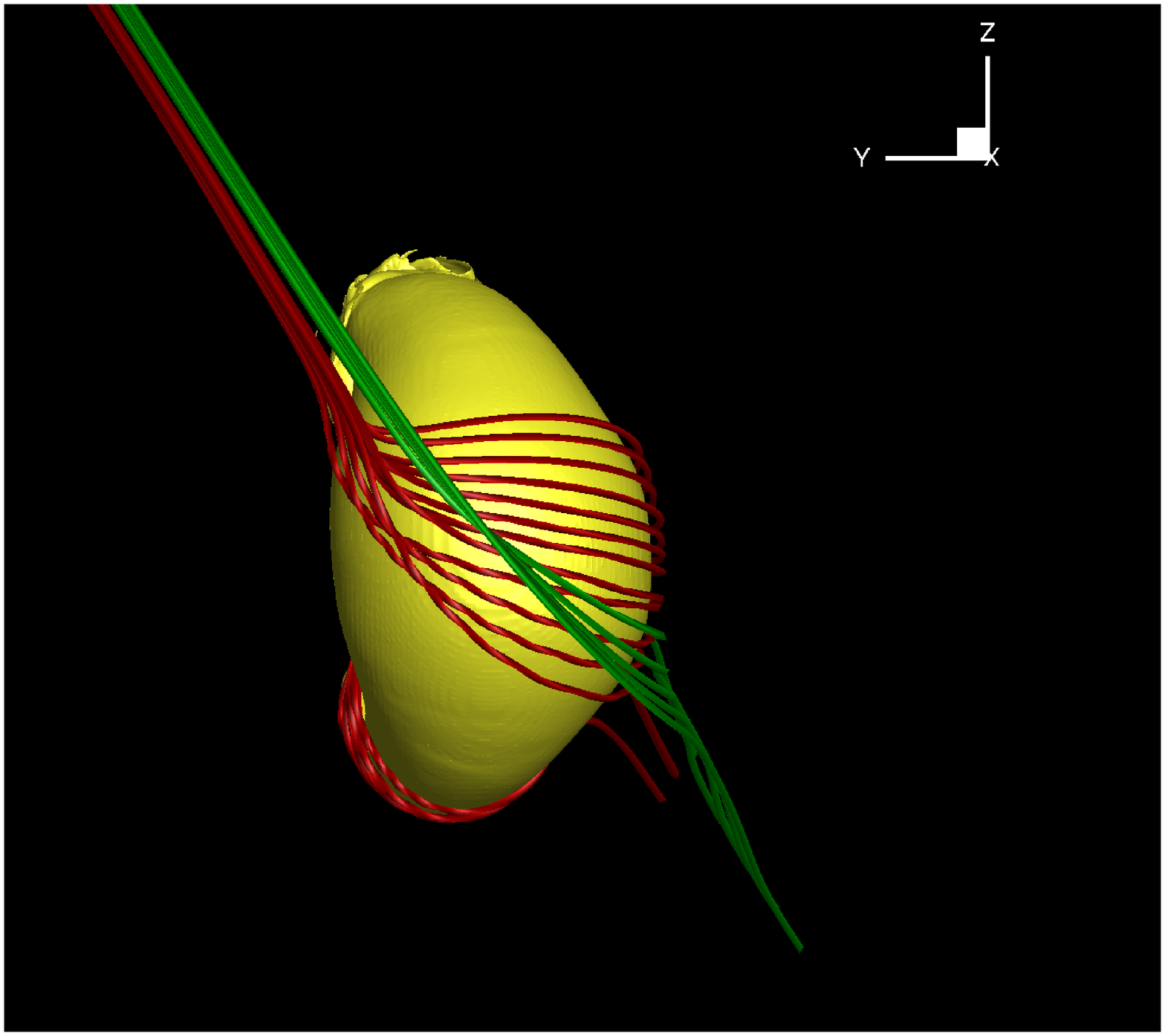}  
\includegraphics[width=0.35\textwidth]{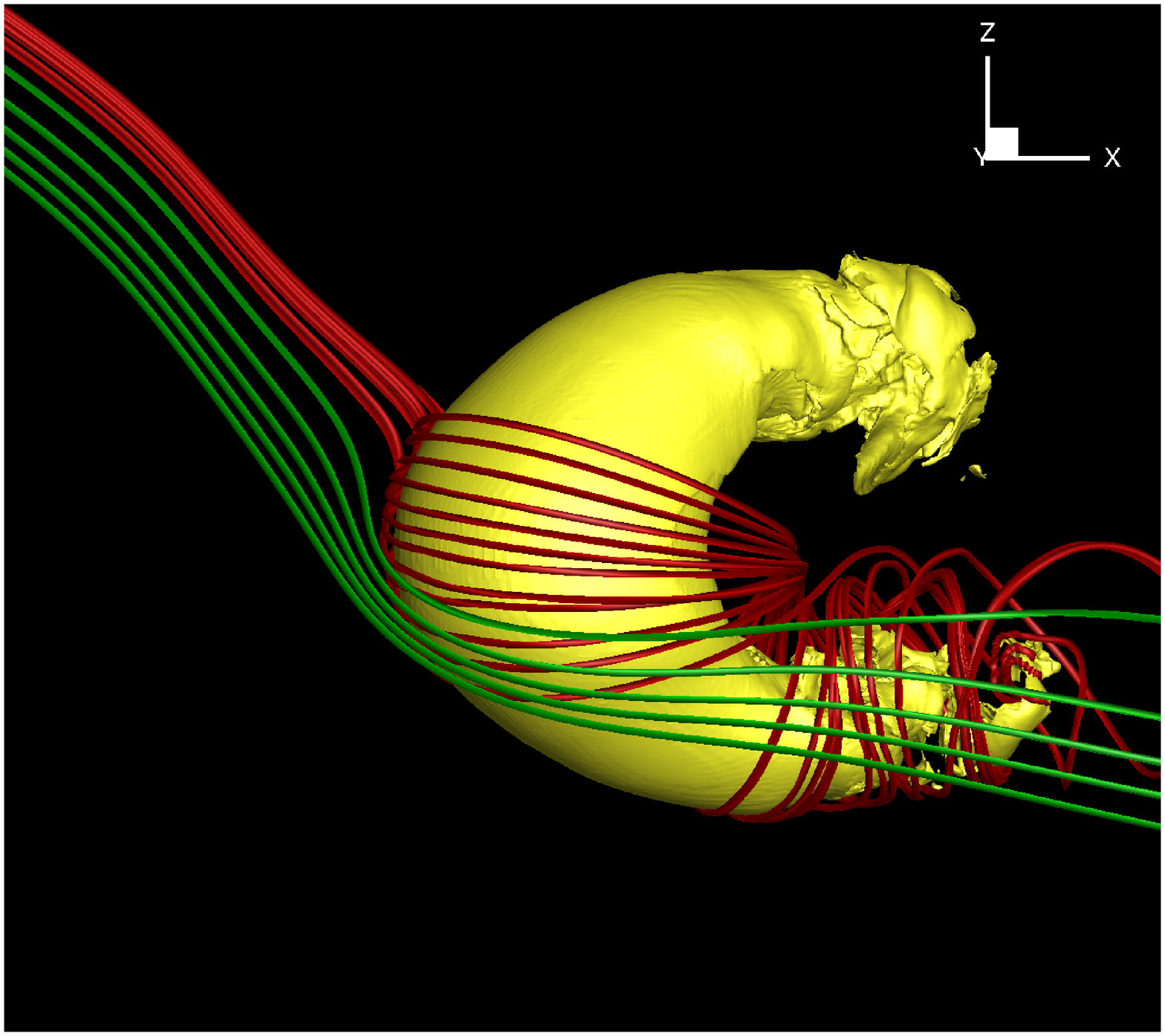}  

\caption{The magnetic field lines ahead of the heliopause. The magnetic field lines just outside of the HP that are reconnected with the solar magnetic field on the eastern flank are shown in red. The unreconnected field lines are shown in green.  The view is from nose in (a) and from the eastern side in (b). The HP is defined with an iso-surface with with $T=2.683 \times 10^{5}$ K. The MHD model used is model A.}
\label{figure5}
\end{figure}

\section{Discussion}
In this paper, we present a model that establishes a physical basis for the twisting of the interstellar magnetic field $B_{ISM}$ into the
direction of the solar magnetic field outside of the HP as measured by Voyager 1. This twist was reported in MHD simulations of the global
heliosphere (Opher \& Drake 2013) but the physics basis for this behavior was not identified. Here we show that the twist of $B_{ISM}$
outside of the HP is a consequence of magnetic reconnection on the eastern flank of the heliosphere, which locally reduces the N
component of $B_{ISM}$ and propagates to the nose as a rotational discontinuity (RD).

What happens if the periodic reversal (due to solar cycle) in the solar magnetic field is included in the model? When the solar field changes sign, reconnection in the western flank will be favored but there should again be an RD upstream of the HP. It seems more likely, however, that because the flows in the HS are slow, a mixture of directions of solar field lines are likely to be present so that reconnection at both flanks is likely to influence the structure of the magnetic field outside of the HP. Such more complex behavior remains to be explored but is unlikely to change the fact that the normal component of $B_{ISM}$ is reduced outside of the HP.

Voyager 1 has been outside the HP for four years and continues to measure an elevation angle $\delta$ that is close to the solar value
in the heliosheath. Based on the new flank reconnection picture, we predict that Voyager 1 will move to a region where $B_{ISM}$ is
causally disconnected (by the Alfv\'enic RD) from the flank reconnection site and the angle $\delta$ will increase to its pristine
interstellar value. 

When a magnetic field line touches the flank and starts to reconnect, the RD will begin propagating upstream. From the simulation we can estimate the time that will take for the RD to propagate upstream. The Alfv\'en speed at the reconnection site is around 
$120~km/s$. Given that the distance of the reconnection site to the nose is around $420$ AU it will take $20$ years for the RD to propagate upstream. 
Estimating the stand-off distance of the RD from the HP is complicated, because the field lines upstream in the simulation are already affected by the draping. 
The stand-off distance will vary depending on the orientation of $B_{ISM}$. In any case, the simulation data suggest that 
$B_{ISM}$ will return to its pristine value $10-15~AU$ past the HP. Similar behavior should eventually be expected at Voyager 2.

Our MHD model, which has led to the flank reconnection picture, is based on the assumption that large-scale reconnection does not take
place in the nose region of the HP. We enforce this in the model by imposing a monopole solar magnetic field with an orientation such
that reconnection does not take place with the interstellar magnetic field at the nose of the HP. In the eastern flank, however, the solar 
magnetic field twists to the N direction and reconnection take
place. Of course, such a model, which was implemented to reduce spurious numerical reconnection in the nose region is not physical.

However, there are solid physics grounds for thinking that reconnection in the flanks of the heliosphere will be much more robust
than in the nose region. It is well-known that MHD simulations do not adequately describe magnetic reconnection, primarily because of the
kinetic length scales that develop during collisionless reconnection (Birn et al 2001). One of the important kinetic effects missed by MHD
models is the stabilizing effect of diamagnetic drifts, which develop at boundaries such as the Earth's magnetopause or the HP
(Swisdak et al., 2003; 2010). The stabilizing influence of these drifts has been extensively documented with solar wind and
magnetospheric satellite data (Phan et al 2010, Phan et al 2013). The diamagnetic drift suppresses reconnection when the drift speed is larger
that the Alfv\'en speed based on the reconnecting magnetic field. The stabilization condition can be written as
\begin{equation}
\Delta \beta > \frac{2L_{p}}{d_{i}} \tan(\frac{\theta}{2})
\end{equation}
where $\Delta\beta$ is the jump in $\beta$ across the HP, $\theta$ is the angle between the magnetic fields on the two sides of the HP and
$L_{p}$ is the typical pressure scale length across the HP, and $d_{i}=c/\omega_{pi}$ is the ion inertial length (with $\omega_{pi}$ the ion plasma frequency). Typically, $L_{p}/d_{i}$ is of order of 1 (Swisdak et al. 2010). Only nearly
anti-parallel reconnection occurs for $\beta \gg 1$. 

In the nose region of the HS, the plasma $\beta$ is large and is dominated by the pickup ions (PUIs) produced in the high-speed solar
wind upstream of the TS. This is in contrast to the low plasma $\beta$ of the interstellar medium. The Voyager 1 spacecraft can not
measure the PUIs, which downstream of the TS have energies greater than a $1$keV. However, estimates based solely on suprathermal tails
($>10keV's$) yield $\beta > 1$ (Krimigis et al. 2010). Therefore, we expect that diamagnetic effects will suppress reconnection in the nose
region of the HP except in localized regions where the magnetic fields across the HP are nearly anti-parallel. This constraint should limit
the scale size of magnetic islands at the HP to a few AU (Swisdak et al 2013, Strumik et al 2014).

The interstellar medium neutrals stream into the heliosphere from the direction of the nose, across the HS and into the high-speed solar
wind, where they charge exchange. In the flanks the flux of neutral atoms across the HS and into the high-speed solar wind is reduced
because the neutral H atoms are moving nearly tangent to the HP and TS. On the flanks the production of PUIs should be greatly reduced
compared with the nose region. The consequence is that the plasma $\beta$ in the HS is much lower in the flanks compared with the nose
and diamagnetic stabilization of reconnection will not take place in the flanks of the HP. Thus, magnetic reconnection will be much more
robust in the flanks than in the nose. While the physics of diamagnetic stabilization is not present in our MHD model the use of
the monopole solar magnetic field, which suppresses reconnection at the nose in the end has the same effect: HP reconnection is suppressed
in the nose and remains robust at in the flanks.

\acknowledgments
The authors would like to thank the staff at NASA Ames Research Center for the use of the Pleiades supercomputer under the award SMD-16-7616. M. O. and B. Z. acknowledge the support of NASA Grand Challenge NNX14AIB0G. J. F. D acknowledges the support of NASA Grand Challenge NNX14AIB0G and NASA award NNX14AF42G. We would like to thank helpful discussions with A. Michael.

\end{document}